\documentclass[]{emulateapj}
\usepackage[]{natbib}
\usepackage{amsmath}
\begin{document}

\title{Random walks and effective optical depth in relativistic flow}   

\author{Sanshiro Shibata\altaffilmark{1}, 
  Nozomu Tominaga\altaffilmark{1,2}, 
  and Masaomi Tanaka\altaffilmark{3}} 
\email{Email: d1221001@center.konan-u.ac.jp}
\altaffiltext{1}{Department of physics, Konan University, 8-9-1 Okamoto, Kobe,
  Hyogo 658-8501, Japan}
\altaffiltext{2}{Kavli Institute for the Physics and Mathematics of the
  Universe, University of Tokyo, 5-1-5 Kashiwanoha, Kashiwa, Chiba 277-8583,
  Japan} 
\altaffiltext{3}{National Astronomical Observatory of Japan, Mitaka, Tokyo
  181-8588, Japan}
\begin{abstract}
We investigate the random walk process in relativistic flow. In the
relativistic flow, photon propagation is concentrated in the directions of the 
flow velocity due to relativistic beaming effect. We show that, in the
pure scattering case, the number of scatterings is proportional to the size
parameter $\xi\equiv L/l_0$ if the flow velocity $\beta\equiv v/c$ satisfies
$\beta/\Gamma\gg \xi^{-1}$, while it is proportional to $\xi^2$ if $\beta/\Gamma\ll
\xi^{-1}$ where $L$ and $l_0$ are the size of the system in the observer frame
and the mean free path in the comoving frame, respectively. We also examine
the photon propagation in the scattering and absorptive medium. We find that,
if the optical depth for absorption $\tau_{\rm a}$ is considerably smaller
than the optical depth for scattering $\tau_{\rm s}$ ($\tau_{\rm
  a}/\tau_{\rm s} \ll 1$) and the flow velocity satisfies $\beta\gg
\sqrt{2\tau_{\rm a}/\tau_{\rm s}}$, the effective optical depth is
approximated by $\tau_*\simeq\tau_{\rm a}(1+\beta)/\beta$.
Furthermore, we perform Monte Carlo 
simulations of radiative transfer and compare the results with the analytic
expression for the number of scattering. The analytic expression is consistent
with the results of the numerical simulations. The expression derived in this
Letter can be used to estimate the photon production site in relativistic
phenomena, e.g., gamma-ray burst and active galactic nuclei. 
\end{abstract}

\section{Introduction}
Relativistic flows or jets are important phenomena in many astrophysical
objects, such as gamma-ray bursts (GRBs) and active galactic nuclei (AGNs).
It is widely accepted that most of high-energy emission from these objects
arises from the relativistic jets. However, their radiation mechanism is not
fully understood. In particular, recent observations of GRBs have indicated
the existence of thermal radiation in the spectrum of the prompt emission,
which casts a question to standard emission models invoking synchrotron
emission.

For example, \citet{2010ApJ...709L.172R} argued that the spectrum of
GRB 090902B can be well fitted by a quasi-blackbody with a characteristic
temperature of $\sim 290~{\rm keV}$. Moreover, it has been reported that some
bursts exhibit a thermal component on a usual non-thermal component
\citep[e.g.,][]{2011ApJ...727L..33G, 2012ApJ...757L..31A}. Therefore,
investigation of the thermal radiation from GRB jets is crucial to understand
the radiation mechanism of GRBs.

The thermal radiation from GRB jets have also been theoretically studied by
several methods as follows: fully analytical studies
\citep[e.g.,][]{2000ApJ...530..292M,2005ApJ...628..847R}, calculations of
photospheric emission which treat the thermal radiation as the superposition
of blackbody radiation from photosphere
\citep[][]{2009ApJ...700L..47L,2011ApJ...732...34L,2011ApJ...732...26M,2011ApJ...731...80N},
and detailed radiative transfer calculations with spherical outflows or
approximate structures of the jets  
\citep[e.g.,][]{2006A&A...457..763G,2012MNRAS.422.3092G,2008ApJ...682..463P,
2010MNRAS.407.1033B,2011ApJ...732...49P, 
  2013MNRAS.428.2430L,2013ApJ...767..139B,2013ApJ...777...62I}. 

To study the thermal radiation, treatment of the photosphere needs careful consideration.
\citet[][]{2009ApJ...700L..47L,2011ApJ...732...34L,2011ApJ...732...26M,2011ApJ...731...80N}
performed the hydrodynamical simulations of relativistic jet and calculated 
the thermal radiation assuming that the photons are emitted at the
photosphere which is defined by the optical depth for electron
scattering $\tau_{\rm s}=1$. However, the observed photons should be produced
in more inner regions with $\tau_{\rm s}\gg 1$
\citep[e.g.,][]{2013ApJ...764..157B} since the radiation and absorption 
processes are very inefficient near the photosphere due to the low plasma
density. The produced photons propagate through the jet and cocoon which have
complicated structure. Thus, radiative transfer calculation of the propagating
photons properly evaluating the photon production site is necessary to investigate
the thermal radiation from GRB jets.

\begin{figure}
  \includegraphics[scale=0.85,clip=true]{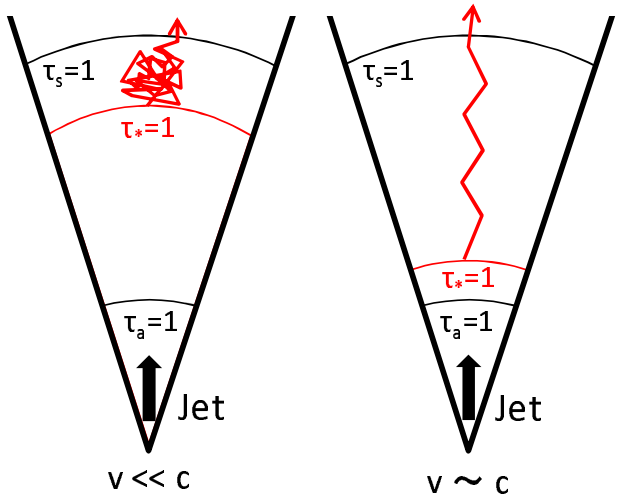}
  \caption{Schematic pictures of photon propagation in the jet with the
    velocity $v\ll c$ (left) and $v\sim c$ (right). When $v\ll c$,
    the scatterings of the photons are approximately isotropic in the observer
    frame and the surface of $\tau_*=1$ is located far from the surface of
    $\tau_{\rm a}=1$, where $\tau_{\rm a}$ is an optical depth in the absence
    of the scatterings. When $v\sim c$, the photons are concentrate
    to the direction of the flow. Thus, the surface of $\tau_*=1$ is
    located close to the surface of $\tau_{\rm a}=1$.
  }
  \label{fig1}
\end{figure}

The photon production site can be estimated by the
effective optical depth $\tau_*$ \citep[e.g.,][]{1979rpa..book.....R}.
However, the expression derived in \cite{1979rpa..book.....R} is based on an 
assumption that each scatterings is isotropic in observer frame. The
assumption does not strictly hold in any moving media because the photon
propagation is concentrated to the direction of the flow due to the beaming effect in the
observer frame (Figure \ref{fig1}).

In this letter, we construct an expression for the effective optical depth 
considering the random walk process in the relativistic flow.
In Section \ref{sec:analytic}, we analytically investigate the random walk
process in relativistic flow and present the expression for the effective
optical depth. In Section \ref{sec:numerical}, we demonstrate that the
number of scatterings obtained by the analytic expression agrees with that
derived by Monte Carlo simulations. Finally, summary and discussions are
presented in Section \ref{sec:summary}.

\section{Analytic expression of random walks in relativistic
  flow}\label{sec:analytic}

In this section, we extend the argument for the random walk process shown
in \cite{1979rpa..book.....R} to the relativistic flow. For simplicity, 
  we assume that the scatterings are isotropic and elastic in the electron
  rest frame.

\subsection{Pure scattering}\label{subsec:randomwalk}
We first consider purely scattering medium with uniform opacity in which
photons are scattered $N$ times. The path of the photons
between $i$-1th and $i$th scattering is denoted by ${\bf r}_{i}$. The net
displacement of the photon after $N$ scatterings is ${\bf R}={\bf r}_1+{\bf
  r}_2+\cdots +{\bf r}_N.$ In order to derive the average net displacement of
photons $l_*$, we first take the square of ${\bf R}$ and then average it,
\begin{equation}\label {lstar2}
  l_*^2\equiv\langle{\bf R}^2\rangle=\sum^N_{i=1}\langle{\bf
    r}^2_i\rangle+\sum^N_{\substack{i,j\\ i\neq j}}\langle{\bf r}_i\cdot{\bf r}_j\rangle,
\end{equation}
where the angle bracket indicates the average for all photons.

If the medium is at rest relative to an observer, the second term in right
hand side of equation (\ref{lstar2}) vanishes due to the front-back symmetry
of the scatterings and only the first term contributes to $l_*$. In this
case, the first term is calculated as $N\langle\overline{{\bf r}^2}\rangle$
where  $\overline{{\bf r}^2}$ is the expected value of the square of the mean
free path. Since the probability that a photon travels a distance
$x$ is $\exp(-x/l)/l$, where $l$ is the mean free path of the photon,
$\overline{{\bf r}^2}$ can be calculated as 
\begin{equation}\label{r2}
\overline{{\bf r}^2}=l^{-1}\int_0^{\infty}x^2\exp(-x/l)dx=2l^2.
\end{equation}
Therefore, since $l$ is the same as the mean free path in the comoving frame
$l_0$ for the static medium and the mean free path is the same for all
photons, $l_*^2=2N\langle l^2 \rangle=2N l_0^2$.\footnote{\label{foot:eq2}This
  is different from the one shown in the \citet{1979rpa..book.....R} by the
  factor of 2. The difference comes from that the first term in
  Eq. (\ref{lstar2}) is estimated approximately as $Nl^2$ in
  \citet{1979rpa..book.....R} but, in this Letter, we calculate the term
  precisely considering the expected value of the square of the mean free path. 
}

The number of scatterings required for a photon to escape a medium which
has a finite width $L_0$ in the comoving frame is
$N=(L_0/l_0)^2/2=\tau_0^2/2$, where $\tau_0$ is the optical depth of the
medium, and this is Lorentz invariant. However, the calculation of the mean
number of scatterings of the photons propagating the distance $L$ in the
observer frame is more complicated because the distances in the two frames
are different and the origin of photon production moves in the observer
frame.

The radius is usually measured in the observer frame especially when one
performs the hydrodynamical simulations and when the emission radius is
observationally measured. Thus, it is useful to construct an
expression in the observer frame to describe the diffusion of
photons. Therefore, we consider mean number of scatterings while the photons
propagate a distance $L$ in the observer frame.

If the medium has a relativistic speed, the second term in right hand side of
Equation (\ref{lstar2}) remains because the photons concentrate in the
velocity directions of the medium due to relativistic beaming
effect. Therefore, the average of scalar products of each path have a
non-zero value in the observer frame. Moreover, the average for the first
term must take into account the dependence on the angle between the directions
of the photon propagation and the flow velocity because the mean free path is
angle dependent in the relativistic flow. Thus, in order to treat the random
walk process in relativistic flow, we need to estimate both $\langle{\bf
  r}^2_i\rangle$ and $\langle{\bf r}_i\cdot{\bf r}_j\rangle$ with taking into
account the relativistic effect.

The mean free path of a photon in the observer frame is given as
$l={l_0}/{\Gamma (1-\beta \cos \theta)}$ \citep{1991ApJ...369..175A},  
where, $\Gamma$, $\beta$, and $\theta$ are fluid Lorentz factor, fluid
velocity in unit of speed of light, and the angle between the directions of
photon propagation and fluid velocity, respectively.
We average $l^2$ integrating in the comoving frame as follows\footnote{The
  integration can also be done in the observer frame with weighting by
  distribution of the photon rays resulted from the beaming effect.}
\begin{equation}
  \langle l^2 \rangle=\frac{l_0^2}{4\pi\Gamma^2}\int_0^{2\pi} d\phi' 
  \int_0^{\pi}\sin\theta' d\theta'(1-\beta\cos\theta)^{-2},
\end{equation}
where the values measured in the comoving frame are denoted with prime.
Using the relation between the angles in the observer frame and the comoving
frame, that is $\cos\theta=(\beta+\cos\theta')/(1+\beta\cos\theta')$, we
obtain  
\begin{equation}\label{r12}
  \langle l^2\rangle=\frac{\Gamma^2(\beta^2+3)}{3}l_0^2,
\end{equation}
and the first term in Equation (\ref{lstar2}) is calculated by $2N\langle
l^2\rangle$.

The scalar product of two paths is ${\bf r}_i\cdot{\bf
  r}_j=l_il_j(\sin\theta_i\cos\phi_i\sin\theta_j\cos\phi_j
+\sin\theta_i\sin\phi_i\sin\theta_j\sin\phi_j+\cos\theta_i\cos\theta_j$).
If we set the polar axis to the direction of the photon propagation,
the azimuthal angle $\phi$ is identical in both frames. Thus only the third
term in the bracket contributes the average and we obtain 
\begin{eqnarray}\label{r1r2}
  \langle{\bf r}_i\cdot{\bf r}_j\rangle&=&\frac{1}{(4\pi)^2}\int d\Omega_i'
  \int d\Omega_j' ~l_i l_j \cos\theta_i \cos\theta_j \nonumber\\ 
  &=&(\Gamma\beta)^2l_0^2.
\end{eqnarray}
Substituting the equation (\ref{r12}) and (\ref{r1r2}) into equation
(\ref{lstar2}), we obtain
\begin{equation}\label{lstar}
  l_*^2=N\frac{2}{3}\Gamma^2(\beta^2+3)l_0^2+N(N-1)(\Gamma\beta)^2l_0^2.
\end{equation}
If we set $l_*=L$, $N$ corresponds to the mean number of scatterings
during the photons propagation of the net distance $L$ in the observer
frame. This leads to a quadratic equation for $N$ as
\begin{equation}\label{eqn}
  (\Gamma\beta)^2N^2+\Gamma^2(2-\frac{\beta^2}{3})N-\xi^2=0,
\end{equation}
where $\xi\equiv L/l_0$ is the size parameter. If the medium is static,
$\xi$ corresponds to the optical depth of the medium. However, in general,
$\xi$ does not correspond to the optical depth because it is defined by the
size of the medium in the observer frame, $L$, and the mean free path of a
photon in the comoving frame, $l_0$.\footnote{
  Since the mean free path in the observer frame depends on the angle between
  the direction of photon propagation and the flow velocity, we define $\xi$
  with $L$ and $l_0$.
}  We employ $\xi$ as the parameter to
parametrize the distance in the observer frame. We can derive $N$ by solving
Equation (\ref{eqn}) as 
\begin{equation}\label{nscat}
  N=\frac{1}{2a}(\sqrt{b^2+4a\xi^2}-b),
\end{equation}
where $a=(\Gamma \beta)^2$ and $b=\Gamma^2(2-{\beta^2}/{3})$.

We derive important indications from equation (\ref{nscat}) as follows:  When
$\xi^2\ll b^2/4a$, which approximately means $\beta/\Gamma \ll \xi^{-1}$, $N$
reduces to $\xi^2/b$ and $N\simeq \xi^2/2$ for non-relativistic
flow\footnote{\label{foot:nscat} This is also different from the one shown in
  the \citet{1979rpa..book.....R}, $N\simeq\xi^2=\tau_0^2$ for the static
  medium, by the factor of 1/2 for the reason argued in Footnote
  \ref{foot:eq2}.}.
However, if $\xi^2\gg b^2/4a$, which means $\beta/\Gamma \gg \xi^{-1}$, 
$N$ becomes $N\simeq \xi/\sqrt{2a}=\xi/\sqrt{2}\Gamma \beta$. 
Thus, when the beaming is effective and the medium is sufficiently optically
thick, $N$ is proportional to $\xi$ with the factor which corresponds to the
reduction of the optical depth for relativistic effect. This is because
photons propagate approximately straight toward the outside and the number of
target electrons during the propagation is proportional to $L\propto\xi$. 

It is noted that the $\xi$ is calculated by $l_0$ which is mean free
path in the comoving frame. Equation (\ref{nscat}) also can be expressed with
 the optical depth $\tau=\Gamma(1-\beta\cos\Theta)\xi$ instead of $\xi$ as
\begin{equation}
  N=\frac{1}{2a}\left(\sqrt{b^2+\frac{4a\tau^2}{\Gamma^2(1-\beta\cos\Theta)^2}}-b\right),
\end{equation}
where $\Theta$ is the angle between the directions along which the optical
depth is measured and the flow velocity.

\subsection{Scattering and absorption}
Next, we consider a photon transfer in a medium involving scattering and 
absorption process. The mean free path of a photon in the
comoving frame is 
\begin{equation}\label{l0}
  l_0=\frac{1}{\alpha_0+\sigma_0},
\end{equation}
where $\alpha_0$ and $\sigma_0$ are absorption and scattering coefficient in the
comoving frame, respectively. 
The probability that a free path ends with a true absorption is
\begin{equation}\label{epsilon}
  \epsilon=\frac{\alpha_0}{\alpha_0+\sigma_0}.
\end{equation}
If we assume that a photon is absorbed after $N$ scatterings, the average
number of scatterings $N$ can be related to the $\epsilon$ by $N=1/\epsilon$.
Substituting this relation and equations (\ref{l0}) and (\ref{epsilon}) into
equation (\ref{lstar}), we obtain $l_*$ as the functions of $\alpha_0$ and
$\sigma_0$:
\begin{equation}
  l_*^2=\left\{\frac{2}{3}\Gamma^2(\beta^2+3)+
  (\Gamma\beta)^2\frac{\sigma_0}{\alpha_0}\right\}\frac{1}{\alpha_0(\alpha_0+\sigma_0)}.
\end{equation}

Introducing the optical depth for absorption and scattering in the observer
frame as $\tau_{\rm a}=\Gamma(1-\beta\cos\Theta)\alpha_0L$ and $\tau_{\rm
  s}=\Gamma(1-\beta\cos\Theta)\sigma_0L$, respectively, the effective optical
depth $\tau_*\equiv L/l_*$ becomes 
\begin{equation}\label{taustar}
  \tau_*=\left\{\frac{2}{3}\Gamma^2(\beta^2+3)+
  (\Gamma\beta)^2\frac{\tau_{\rm s}}{\tau_{\rm a}}\right\}^{-1/2}
\frac{\sqrt{\tau_{\rm a}(\tau_{\rm a}+\tau_{\rm s})}}{\Gamma(1-\beta\cos\Theta)}.
\end{equation}

In the non-relativistic limit, equation (\ref{taustar}) reduces to
$\tau_*=\sqrt{\tau_{\rm a}(\tau_{\rm a}+\tau_{\rm s})/2}$, which is consistent with
the effective optical depth in the static medium shown in
\citet{1979rpa..book.....R} except for the factor of $1/\sqrt{2}$
 (see Footnotes \ref{foot:eq2} and \ref{foot:nscat}).

\begin{deluxetable}{l||c|c|c}
  \tablewidth{0pt}
  \tablecaption{Approximate forms of effective optical depth $\tau_*$}
  \tablehead{
    $\beta$ & $\beta\ll\sqrt{2\tau_{\rm a}/\tau_{\rm s}}$ & $\sqrt{2\tau_{\rm a}/\tau_{\rm s}} \ll \beta \ll 1 $& $\beta\sim 1$
  }
  \startdata
  $\tau_*$ & $\sqrt{\tau_{\rm a}\tau_{\rm s}/2}$ & $\tau_{\rm a}/\beta$ & $2\tau_{\rm a}$

  \tablecomments{ The top and bottom lines represent the ranges of the velocity $\beta$
    and approximated forms of effective optical depth $\tau_*$ in the ranges of
    $\beta$, respectively.}
  \enddata
  \label{tab:taustar}
\end{deluxetable}

Here, we consider scattering dominant case, i.e., $\tau_{\rm s}\gg\tau_{\rm a}$,
which is the case in the GRB jets and cocoon.
In this case, the behavior of $\tau_*$ depends on the relation between $\beta$
and $\tau_{\rm a}/\tau_{\rm s}$. If $\beta\ll \sqrt{2\tau_{\rm a}/\tau_{\rm s}}$ ($\ll 1$), $\tau_*$
becomes $\tau_*\simeq \sqrt{\tau_{\rm a}\tau_{\rm s}/2}$. On the other hand,
if $\beta\gg \sqrt{2\tau_{\rm a}/\tau_{\rm s}}$, $\tau_*$ is approximated by 
\begin{equation}\label{taustar2}
  \tau_*\simeq \frac{\tau_{\rm a}}{\Gamma^2\beta(1-\beta\cos\Theta)}.
\end{equation}
If we calculate the optical depth along the velocity direction, i.e.,
$\Theta=0$,
\begin{equation}
  \tau_*\simeq\frac{1+\beta}{\beta}\tau_{\rm a}.
\end{equation}
This can be approximated as
\begin{equation}\label{taustar3}
  \tau_*\simeq \frac{\tau_{\rm a}}{\beta} \gg \tau_{\rm a},
\end{equation}
for the non-relativistic flow and
\begin{equation}
  \tau_*\simeq 2\tau_{\rm a},
\end{equation}
for the relativistic flow.  Therefore, the dependence of $\tau_*$ on $\tau_{\rm
  a}$ is different for $\beta\ll\sqrt{2\tau_{\rm a}/\tau_{\rm s}}$ and
$\beta\gg\sqrt{2\tau_{\rm a}/\tau_{\rm s}}$. The effective optical depth
$\tau_*$ is proportional to $\tau_{\rm a}$ when $\beta\gg\sqrt{2\tau_{\rm
    a}/\tau_{\rm s}}$ for the same reason that the number of scatterings is
proportional to $\xi$ when $\beta/\Gamma\gg \xi^{-1}$ in the pure scattering
case as argued in Section \ref{subsec:randomwalk}.
 We summarize these approximated forms of $\tau_*$ for various ranges of
 $\beta$ in Table \ref{tab:taustar}.

The effective optical depth defines the photon production site as $\tau_*=1$.
From Equation (\ref{taustar3}), $\tau_*$ is much larger than $\tau_{\rm a}=1$
as long as $\beta\ll1$ even for $\beta\gg\sqrt{2\tau_{\rm a}/\tau_{\rm
    s}}$. This indicates that the photon production site in the flow with
$\beta\ll1$ is located at much outer region than the surface of $\tau_{\rm a}=1$
as illustrated at the left of Figure \ref{fig1}. On the other hand, when the
flow has relativistic velocity, $\tau_*$ differs from $\tau_{\rm a}$ by only
the factor of 2 and the photon production site is located close to the surface
of $\tau_{\rm a}=1$ as illustrated at the right of Figure \ref{fig1}.

It should be noted that, even if the flow is non-relativistic, $\tau_*$
departures from the one for the static medium as long as the conditions of
$\tau_{\rm s}\gg\tau_{\rm  a} $ and $\beta\gg\sqrt{2\tau_{\rm a}/\tau_{\rm
    s}}$ are satisfied. This is because that a large number of scatterings
makes the effect apparent even if the relativistic beaming has only a small
effect at each scattering.

\section{Monte Carlo simulations}\label{sec:numerical}
\subsection{Numerical code}
We developed a radiative transfer code based on Monte Carlo
method. Only the Compton scattering process is taken into account and any
3-dimensional structures of density and temperature can be treated.

The probability $P_{\rm s}$ that a photon scatters on an optical depth
$\delta\tau$ is estimated as $P_{\rm s}=1-\exp(-\delta\tau)$. $\delta\tau$ is
written with a distance $\delta s$ as $\delta\tau=\Gamma_{\rm e}(1-\beta_{\rm
  e}\cos\theta)n_{\rm   e}\sigma_{\rm KN} \delta s$, where $\Gamma_{\rm
  e}$, $\beta_{\rm e}$, $n_{\rm e}$ are the electron Lorentz factor, the
electron velocity in unit of speed of light in the observer frame, and the
electron number density in the comoving frame, respectively. Contributions
from both the fluid bulk motion and the thermal motion of electrons are taken
into account in $\Gamma_{\rm e}$ and $\beta_{\rm e}$. The $\sigma_{\rm KN}$ is the
Klein-Nishina cross section. The occurrence of scattering during the travels
of $\delta s$ is evaluated with a uniform random number $R_1$ with a range of
0 to 1. If $R_1 > P_{\rm s}$, the scattering does not occur and the photon
freely travels the distance $\delta s$. If $R_1<P_{\rm s}$, the photon is
scattered by an electron at a distance $l_{\rm s}<\delta s$ which is
calculated with the $R_1$ as
\begin{equation}
  l_{\rm s}=\frac{-\ln(1-R_1)}{\Gamma_{\rm e}(1-\beta_{\rm e}\cos\theta)n_{\rm e}\sigma_{\rm
      KN}}.
\end{equation}

The thermal motion of electrons in the fluid comoving frame follows the
relativistic Maxwell distribution function $f(p)\propto p^2\exp(-\sqrt{m_{\rm
    e}c^4 + p^2c^2}/kT)$, where $p$ is the momentum of the electrons about the
thermal motion \citep[e.g.,][]{1980stph.book.....L}. We assume that the
electrons move isotropically in the fluid comoving frame.
  
The scatterings alter the energy and the direction of the photon. We calculate
the four-momentum of the photon during the scattering as
follows: the four-momentum of the photon before scattering is Lorentz
transformed to the electron rest frame, the photon scatters with Klein-Nishina
cross section, and then the four-momentum after the scattering is Lorentz
transformed into the observer frame.

\subsection{Results}
In order to confirm the analytic arguments in Section \ref{sec:analytic}, we
perform radiative transfer simulations with the Monte Carlo method for the
photons scattered in the relativistic flow. We compare the mean number of 
scatterings $\langle N\rangle$ with Equation (\ref{nscat}).

We consider a uniform flow with a velocity $\beta$ and the electron
number density $n_{\rm e}$ of $10^{-10}/\sigma_{\rm T} \mathrm{~cm^{-3}}$,
where $\sigma_{\rm T}$ is Thomson scattering cross section. The flow
velocity is parallel to $z$ direction. Photons are created at the
origin of the coordinate with an energy of $E_{\rm ph}=0.1~\mathrm{eV}$,
which is set as to avoid the Klein-Nishina effect, in the comoving frame.
We calculate the mean number of scatterings $\langle N\rangle$ while the 
photons travel a net distance $L$ which ranges from $10^{11}$ to
$10^{14}~\mathrm{cm}$ in the observer frame, so that the corresponding
$\xi=n_{\rm   e}\sigma_{\rm T}L$ ranges from $10$ to $10^4$.

Since our interests in this Letter is the influence of the fluid bulk motion
on the number of scatterings, the temperature of the medium is set to be very
low, i.e., $kT=1\mathrm{eV}$, to avoid that the thermal motion of electrons
affect the number of scatterings. We investigate non-relativistic and
relativistic velocity of the medium with products of the Lorentz factor
and the velocity $\Gamma\beta$ of $10^{-3}$, $10^{-2}$, $10^{-1}$, $1,$ $10$,
and $10^2$.

\begin{figure}
\plotone{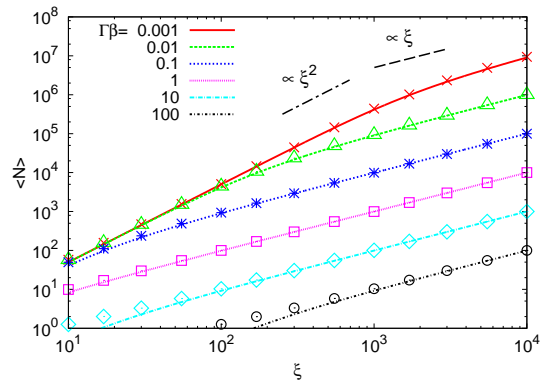}
\caption{The mean number of scatterings $\langle N\rangle$ for the models
  with $\Gamma\beta=10^{-3}$ (crosses), $10^{-2}$ (triangles), $10^{-1}$
  (asterisks), 1 (squares), $10$ (diamonds), and $100$ (circles). The lines
  show the analytic expressions, equation (\ref{nscat}), for the models
  denoted in the figure.
}  
\label{fig2}
\end{figure}

Figure \ref{fig2} shows the mean number of scatterings
$\langle N\rangle$ of $6\times 10^3$ photons for the models with 
$\Gamma\beta=10^{-3}$, $10^{-2}$, $10^{-1}$, 1 and $6\times 10^4$ photons
for the models with $\Gamma\beta=10$ and $100$. The lines show the analytic
expressions derived in the previous section with $\Gamma\beta=10^{-3},
10^{-2}, 10^{-1}, 1, 10$ and $100$ (Eq. (\ref{nscat})). This demonstrates that
the analytic expressions are excellently consistent with the results of
numerical simulations, except at $\langle N\rangle \sim 1$. The
difference at the region comes from the fact that a
considerable number of photons do not experience any scatterings in this
region, although the equation (\ref{nscat}) is obtained assuming all the
photons undergo more than one scatterings.

The dependencies of $\langle N\rangle$ on $\xi$ are as follows:
In the model with $\Gamma\beta=10^{-3}$, $\langle N\rangle$ is proportional
to $\xi^2$ for $\xi<10^3\simeq \beta^{-1}$ and to $\xi$ for $\xi>10^3$.
In the model with $\Gamma\beta=10^{-2}$, $\langle N\rangle$ is proportional to
$\xi^2$ for $\xi<10^2$ and to $\xi$ for $\xi>10^2$. The transition of the
dependence is at $\xi\sim\beta^{-1}$. In the models with
$\Gamma\beta=10^{-1}$, 1, and $10$, $\langle N\rangle$ is proportional
to $\xi$ in the range of $10<\xi<10^4$.
In the model with $\Gamma\beta=100$, $\langle N\rangle$ is proportional to
$\xi$ for $\xi > 10^2$.

\section{Summary \& discussions}\label{sec:summary}
In this letter, we investigate the random walk process in relativistic flow.
In the pure scattering medium, the mean number of scatterings at the
size parameter of $\xi$ is proportional to $\xi^2$ for $ \beta/\Gamma\ll
\xi^{-1}$ and to $\xi$ for $\beta/\Gamma\gg \xi^{-1}$. These dependencies of
the mean number of scatterings on $\xi$ are well reproduced by the numerical
simulations. We also consider the combined scattering and absorption case. If
the scattering opacity dominates the absorption opacity, the behavior of the
effective optical depth is different depending on the velocity $\beta$. If
$\beta\ll \sqrt{2\tau_{\rm a}/\tau_{\rm s}}$, the effective optical depth is 
$\tau_*\simeq \sqrt{\tau_{\rm a}\tau_{\rm s}/2}$ and if $\beta\gg
\sqrt{2\tau_{\rm a}/\tau_{\rm s}}$, $\tau_*\simeq (1+\beta)\tau_{\rm a}/\beta$.

In the GRB jets, the flow has ultra-relativistic velocity ($\Gamma\gtrsim100$) 
and the electron scattering opacity dominates the absorption opacity
($\tau_{\rm s}\gg\tau_{\rm a}$) due to its low density and high
temperature. Thus, the effective optical depth in the jet is approximated by 
$\tau_{*}\simeq 2\tau_{\rm a}$. On the other hand, the cocoon have a
non-relativistic velocity \citep[e.g.,][]{2003MNRAS.345..575M} and the
effective optical depth in the cocoon could be much higher than the absorption
optical depth as $\tau_*\simeq\tau_{\rm a}/\beta\gg \tau_{\rm a}$. The
effective optical depth defines the photon production site as $\tau_*=1$. In
the subsequent papers, we will perform the radiative transfer calculations for
the thermal radiation from GRB jet and cocoon taking into account the photon
production at the surface of $\tau_*=1$. This enables us to correctly treat the
photon number density at the photon production sites.

The results could be applicable not only for GRB jet and cocoon but also for
the other astronomical objects such as AGNs or black hole binaries. For
example, the super critical accretion flows around the black holes produce a
high temperature ($\sim10^8~\mathrm{K}$) and low density ($\sim
10^{-9}~\mathrm{g/cm^3}$) outflow with a semi-relativistic velocity ($\sim
0.1c$) \citep[e.g.,][]{2009PASJ...61..769K}. In these circumstances, the
scattering process have a major role on the photon diffusion and the
relativistically corrected treatment is necessary even though the flow
velocity is rather small compared with the speed of light.

\acknowledgements
This research has been supported in part by
World Premier International Research Center Initiative, MEXT, Japan, by
the Grant-in-Aid for Scientific Research of the JSPS (23740157, 24740117,
25$\cdot$2912) and by the Grant-in-Aid for Scientific Research on Innovative
Areas of the MEXT (25103515).

\end{document}